\documentclass[journal]{IEEEtran}

\usepackage{amssymb}
\usepackage{stmaryrd}
\usepackage{subfig}
\usepackage {rotating,cite,subeqn, graphicx, calc,color,eepic,marvosym,pxfonts,txfonts}
\usepackage{ifsym}
\usepackage[font = footnotesize]{caption}
\usepackage{authblk}
{}
{}
{}

\title{Orthogonal Frequency Division Multiplexed Quantum Key Distribution}
\author[1]{Sima Bahrani,~\IEEEmembership{Student Member,~IEEE}, Mohsen Razavi, and Jawad A. Salehi,~\IEEEmembership{Fellow,~IEEE}
\thanks{This research was supported in part by the European Community's Seventh Framework Programme Grant Agreement 277110, the UK's Engineering and Physical Sciences Research Council Grant EP/M506953/1, and Iran National Science Foundation (INSF).}
\thanks{S. Bahrani and J. A. Salehi are with the Department of Electrical Engineering, Sharif University of Technology, Tehran, Iran. M. Razavi is with the School
of Electronic and Electrical Engineering, University of Leeds, Leeds, LS2 9JT,
UK, e-mail: m.razavi@leeds.ac.uk.}
}

\begin{document}
\maketitle
\begin{abstract}
We propose orthogonal frequency division multiplexing (OFDM), as a spectrally efficient multiplexing technique, for quantum key distribution (QKD) at the core of trusted-node quantum networks. Two main schemes are proposed and analyzed in detail, considering system imperfections, specifically, time misalignment issues. It turns out that while multiple service providers can share the network infrastructure using the proposed multiplexing techniques, no gain in the total secret key generation { rate} is obtained if one uses conventional passive all-optical OFDM decoders. To achieve a linear increase in the key rate with the number of channels, an alternative active setup for OFDM decoding is proposed, which employs an optical switch in addition to conventional passive circuits. { We show that by using our proposed decoder the bandwidth utilization is considerably improved as compared to conventional wavelength division multiplexing techniques.}  
\end{abstract}

\IEEEpeerreviewmaketitle 

\section{Introduction}
\IEEEPARstart{Q}{uantum} communications has entered a new phase in its development targeting new markets and aiming at widespread use and adoption in different scenarios. With the successful demonstration of SECOQC \cite{secoqc} and Tokyo \cite{Sasaki:TokyoQKD:2011} quantum key distribution (QKD) networks, we are now at a stage to develop many-user quantum networks { \cite{Razavi_MulipleAccessQKD, townsend1997quantum, kitayama2011security, ciurana2014quantum}}. The reach of conventional QKD links is, nevertheless, limited as they rely on low-power signals, e.g., single photons \cite{BB_84}. The initial solution perceived for the first generation of quantum networks relies on a {\em trusted} set of nodes, in a mesh topology, at the core network. Such nodes enable secure key exchange between any two remote users via a cascade of key exchanges between neighboring nodes along the path that connects the two users. In order to support many users at the access nodes, it is necessary to proportionally generate longer secret keys between the internal core nodes of the network. The analogy in classical telecommunications is the ratio between the end-user data rates and the high traffic of data at the backbone of the network. One simple idea to achieve higher key rates is to use multiplexing techniques to generate keys in parallel. In this paper, we employ one of the most advanced classical multiplexing techniques to come up with orthogonal frequency division multiplexed QKD (OFDM-QKD) schemes. We look at existing {\em all-optical} orthogonal frequency division multiplexing (OFDM) techniques \cite{hillerkuss2010simple, hillerkuss201126, wang2011optical, shimizu2012demonstration, schroder2014all, chen2009all} and partly modify their setups in order to obtain spectrally efficient high-rate OFDM-QKD schemes.

QKD enables secure key exchange without relying on computational complexity. This is in contrast with existing techniques for key exchange, e.g., the RSA protocol \cite{rivest1978method}, whose security is at risk with the advancement of technology \cite{ShorAlgorithm_94}. In that sense, QKD provides a {\em future-proof} method of secure communications. The first proposed QKD protocol by Bennett and Brassard in 1984 (BB84) \cite{BB_84} relied on the polarization encoding of single photons. Since then new protocols and encoding schemes have emerged and QKD has seen field demonstrations along with conventional telecom channels \cite{Sasaki:TokyoQKD:2011, Stucki:Long:2011, WDM_QKD_Han_2009, Shields.PRX.coexist}. Recent demonstrations cover distances over 250~km \cite{Wang:260kmQKD:2012} and with nearly 50 users. The next step for QKD development will focus on extending the reach of the system and the number of users QKD networks can support.  

Quantum networks are facing several challenges before their full implementation. One key requirement is their integration with existing and future classical optical communication networks {\cite{maeda2009technologies,ciurana2014quantum}}. This implies the need for new quantum friendly standards for optical networks. That will include devising proper mechanisms by which weak quantum signals can be separated from classical channels \cite{Shields.PRX.coexist, Townsend_QI_home_2011}. Multiple-access techniques are also needed to enable interference-free access to different quantum users \cite{Razavi_MulipleAccessQKD, Razavi_IWCIT12}. Eventually, QKD systems must improve their performance in terms of rate-versus-distance behavior and cost.

\begin{figure}[!t]
\centering
\includegraphics[width=8.6cm]{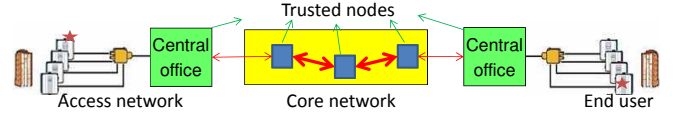}
\caption{Trusted-node architecture for emerging quantum networks. The end users may be connected to the core network via passive optical networks. In order to generate a secret key between two end users, one must first generate a key between any two neighboring nodes along their connecting path. The key generated between the end user and its corresponding central office can then be encrypted and securely relayed node by node until it reaches the other party. Note that the internal links (thicker lines) must carry a higher traffic.
\label{Fig:trustednode}}
\end{figure}

One feasible approach to long-distance QKD is based on trusted-node quantum networks. With current technology, we are able to generate secret keys at a rate on the order of Mb/s at 50~km of distance \cite{Shields.PRX.coexist}. By cascading several of such links, as shown in Fig.~\ref{Fig:trustednode}, {\em and} trusting all intermediate nodes, one, in principle, can exchange secret keys at any distance by first generating secret keys between neighboring nodes and then relaying the initial key, in an encrypted way, to the other party. The main requirement for this approach is to trust all nodes in between the two end users. While this assumption may be acceptable for the first generation of quantum networks, it can be removed in future generations by relying on quantum repeater setups at the core network \cite{Nicolo_paper3}.
  
In trusted-node networks, the internal nodes in the core network are expected to have a high traffic of key exchange as they are providing service to a large number of end users. It is important then to generate a large number of secret key bits per allocated wavelength to each quantum channel over these core links. One possible approach is to use non-binary signalings to send more key bits per transmitted quantum state \cite{sheridan2010security}. In addition to this, we should think how most efficiently we can use the available bandwidth per allocated wavelength, { especially with reference to QKD systems that rely on dense wavelength division multiplexing (DWDM) techniques \cite{ciurana2014quantum}}. Our proposed solution here relies on one of the most spectrally efficient methods in classical communications, i.e., OFDM. OFDM relies on the full orthogonality of its subcarriers to multiplex multiple channels. This full orthogonality is essential in QKD applications \cite{Razavi_MulipleAccessQKD, Razavi_QCMC_2010}, in order to minimize the interference from other classical and quantum users. In our case, each subcarrier represents a QKD channel between two core nodes. The total key generation rate between these two nodes is then expected to increase linearly with the number of subcarriers. Moreover, OFDM is compatible with non-binary signaling techniques, and that would enable us to take the maximum benefit from the available bandwidth. Finally, by using a multiplexing technique, multiple service providers can use the capacity of the core network without trusting each other. Note that the OFDM-QKD can be modified to be used as a multiple-access technique in multi-user QKD setups. In this paper, we focus on the multiplexing aspect with the objective of increasing the rate at the core of QKD networks.

Being an optical system, QKD can be merged best with OFDM if all-optical OFDM encoders and decoders are used. Here, we consider two possible implementations for the all-optical OFDM transmitter. In the first approach, the OFDM subcarriers are generated directly by a bank of frequency offset locked laser sources or an optical comb generator \cite{hillerkuss2010simple, hillerkuss201126, wang2011optical}. After encoding the subcarriers, an optical coupler combines them to generate the OFDM signal. The second approach uses the optical inverse discrete Fourier transform (OIDFT) circuit to generate the OFDM signal \cite{wang2011optical, shimizu2012demonstration, schroder2014all, chen2009all}. Short pulses are fed into the OIDFT circuit following the QKD encoding stage. Both these approaches rely on real-time optical discrete Fourier transform (ODFT) at their receivers. Conventional passive implementations of ODFT turn out to be too lossy to be useful for our main objective of increasing the rate. In our work, we show how the ODFT circuit can be modified to be effective for QKD applications. 

{Different QKD protocols can be used in our proposed OFDM-QKD setups. Here, we focus on the decoy-state variant of the BB84 protocol \cite{Lo:Decoy:2005}. The decoy-state technique allows us to use {\em weak} laser pulses, rather than ideal single-photon sources as originally proposed in \cite{BB_84}, and that would simplify the encoding equipment of QKD. In order to obtain immunity against the photon-number splitting attacks, in the decoy-state protocol, for every transmitted QKD pulse, the sender has to randomly choose its intensity from a set of available intensities, where one of which corresponds to the main signal, and the rest to decoy states. In practice, it is often sufficient to use only two decoy states \cite{MXF:Practical:2005}, although, in this paper, for analytical convenience, we assume infinitely many decoy states are used. Our proposed setups are compatible with other QKD protocols, such as continuous-variable or distributed-phase QKD protocols \cite{Scarani:QKDrev:2008}. The detailed analysis of the latter systems is, however, beyond the scope of this paper.}

 { Both optical OFDM and QKD are advanced technologies. It is interesting to see how drawbacks in one system would translate into the other. While some of the drawbacks with OFDM may directly affect our OFDM-QKD system, there are certain issues that are less of a problem in a QKD setup. For instance, one known OFDM problem in the classical domain is its high peak-to-average power ratio, which makes it susceptible to distortions due to nonlinearity effects\cite{armstrong2009ofdm,shieh2009ofdm}. Fortunately, for QKD applications, nonlinearity is not necessarily a major issue because the QKD transmitted signals are low power. Nevertheless, the common OFDM-related imperfections such as time misalignment \cite{chandrasekhar2009experimental,ali201516,ali2015time}, phase noise introduced by the lasers \cite{shieh2009ofdm,pollet1995ber,colavolpe2011impact}, and frequency offsets between the transmitter and the receiver carriers (in the case of applying a local oscillator at the receiver) \cite{shieh2009ofdm,pollet1995ber} can potentially influence the orthogonality between subchannels, and subsequently affect the performance of OFDM-QKD systems. In this paper, we specifically consider the degrading effects due to time misalignmnet, which is the major source of error in the most promising setup we propose here. 

Finally, it is interesting to note that, while one of the key advantages of OFDM in the microwave domain is its reliance on digital signal processing, OFDM-QKD setups may less benefit from this feature. At the receiver side, any measurement on the OFDM signal before the QKD decoders could alter the transmitted states and result in errors. That is why it is important to have a fully optical setup for OFDM decoders. At the transmitter side, an optical OFDM setup would, in principle, allow multiple users to encode their key bits without trusting each other. This cannot necessarily be achieved if we first generate the OFDM signal electronically and then convert it to an optical signal. That said, there will be much room for improvement in future work, while, in this work, we assess the possibility of OFDM-QKD systems.}

The rest of the paper is organized as follows. In the next section, we propose two OFDM-QKD schemes and describe their principles of operation. In Sec. III, the proposed OFDM-QKD schemes are analyzed from a quantum mechanical perspective. The analysis of secret key generation rate is presented in Sec.~IV, in which we particularly focus on time misalignment issues within the OFDM system. We propose an optimal gating solution to maximize the key rate. Some numerical results are then presented in Sec.~V. We conclude the paper in Sec.~VI.

\section{System Description} 
\label{Sec:Description}
\begin{figure}[!t]
\centering
\includegraphics[width=\linewidth]{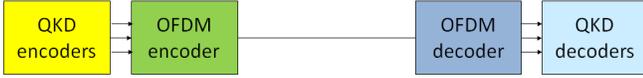}
\caption{QKD over an OFDM link. The QKD encoded optical pulses are multiplexed by an all-optical OFDM encoder. At the receiver, the corresponding OFDM decoder followed by QKD decoders are used to generate secret keys. \label{Fig:genScheme}}
\end{figure}

In this section, we describe QKD over all-optical OFDM links. Figure~\ref{Fig:genScheme} shows the overall system structure. The QKD encoders generate the quantum signals, in the form of pulses, that carry the information about the encoded key bits in each subchannel. The resulting optical pulses are fed simultaneously into the OFDM encoder to be multiplexed. At the receiver, the OFDM decoder followed by essential QKD decoding modules can be used to complete the QKD protocol. The key part in the OFDM decoder is an ODFT circuit, which effectively separates the subchannels. 

\begin{figure}[!b]
\centering
\includegraphics[width=\linewidth]{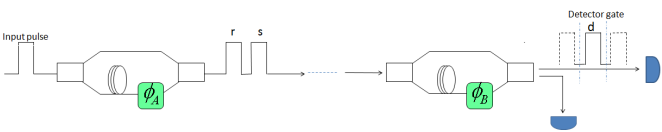}
\caption{Phase encoded QKD. Alice encodes her key bits by choosing a phase value $\phi _A \in \{0,\pi /2,\pi, 3 \pi /2\}$. Each optical pulse passes through the MZI and produces two output pulses with the relative phase $\phi _A$. On the Bob's side, a similar MZI is used to recombine $r$ and $s$ modes, followed by photodetection. \label{Fig:phase_enc}}
\end{figure}

In this paper, we assume that the QKD encoders perform phase encoding using decoy-state techniques \cite{Lo:Decoy:2005}. Based on the phase-encoded BB84 protocol, Alice chooses her phase value $\phi _A$ from one of the bases $\{0,\pi\}$ or $\{\pi /2, 3 \pi /2\}$. The two phase values in each basis correspond to bits 0 and 1. As shown in Fig.~\ref{Fig:phase_enc}, an optical pulse sent by Alice passes through a Mach-Zehnder interferometer (MZI). The output is two non-overlapping successive pulses, denoted by $r$ and $s$, of duration $T$ with a relative phase corresponding to the chosen basis and the transmitted key bit. The QKD decoding includes Bob's selection of his measurement basis by choosing the phase $\phi_B \in \{0,\pi /2 \}$ randomly in one arm of his MZI and the detection of the output signal. In the following, we describe two OFDM-QKD setups based on the proposed schemes for all-optical OFDM.

\subsection{Scheme I}
\label{Sec:Scheme1}
 
\begin{figure}[!t]
\centering
\includegraphics[width=\linewidth]{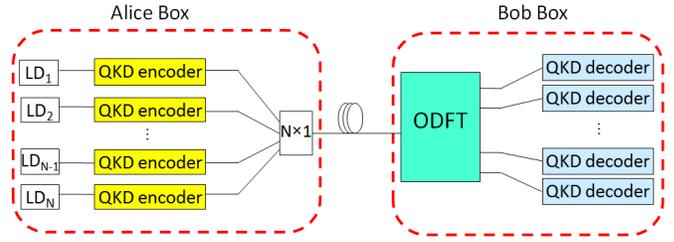}
\caption{OFDM-QKD using directly generated subcarriers. The optical pulses, generated by $N$ frequency offset-locked laser diodes, are fed into the QKD encoders. At the receiver, an ODFT circuit is required to separate the subcarriers. \label{Fig:scheme1}}
\end{figure}

Figure~\ref{Fig:scheme1} depicts the OFDM-QKD system that relies on directly generated subcarriers. At the transmitter, a bank of frequency offset locked laser diodes generate the input optical pulses to $N$ QKD encoders. { These pulses are individually phase randomized, as required by the decoy-state protocol \cite{Random_phase_attack}, and then go through a bank of encoders as in Fig.~\ref{Fig:phase_enc}. Because the information is encoded in the phase difference, these overall random phases do not change the encoded bits. The same holds for the possible phase noise of the lasers so long as their phase is constant during the transmission of each bit. In our forthcoming analysis, we account for possible relative phase distortions between $r$ and $s$ pulses in Fig.~\ref{Fig:phase_enc}.} The outputs of the QKD encoders are then combined to form the OFDM signal. { If we trust all the elements in the Alice box of Fig.~\ref{Fig:scheme1}, we can adjust the transmitted power such that it is at the {\em output} of the combiner that each subchannel has the right intensity for its corresponding pulse. This will allow us to neglect the losses in the encoder box, as we assume in this paper.} At the receiver, ODFT is used to demultiplex the subcarriers. To comply with the OFDM orthogonality condition, in Scheme I, the pulse width is $T=1/{\Delta f}$, where ${\Delta f}$ is the frequency separation of the subcarriers.  

To illustrate the principles of this scheme, consider the classical case, where the OFDM signal is generated by combining classical subchannels as follows:
\begin{equation}
{x} (t)=\sum_{k=0}^{N-1} {{a}_{k} e^{j\omega _{k} t}},\;\;\;\; 0 < t < T,
\label{scheme1_signal}
\end{equation}
where $a_{k}$ is the complex amplitude of the $k^{th}$ subchannel with frequency $\omega_k = \omega_0 + 2 \pi k \Delta f$ for a nominal channel frequency $\omega_0$. { The ODFT circuit, at the decoder, will then separate different subcarriers and generate the following output signals:}
\begin{equation}
y_{m}(t)=\frac{1}{N}\sum_{n=0}^{N-1}{x(t-n T_c)e^{j2\pi nm/N}}, \;\; m=0,1,...,N-1,
\label{scheme1_DFT}
\end{equation}
where $T_c \triangleq T/N$. With the assumption of $T=1/{\Delta f}$, we can conclude from (\ref{scheme1_signal}) and (\ref{scheme1_DFT}) that
\begin{equation}
{y}_m (t)=\sum_{k=0}^{N-1}{{a}_{k} e^{j\omega_{k}t} ( \frac{1}{N}\sum_{n=0}^{N-1}{e^{j2\pi n (m-k)/N}})}.
\end{equation}
The term in the brackets is nonzero only if $k=m$, which leads to the $m^{th}$ subcarrier extraction. 

\begin{figure}[!t]
\centering
\includegraphics[width=3in]{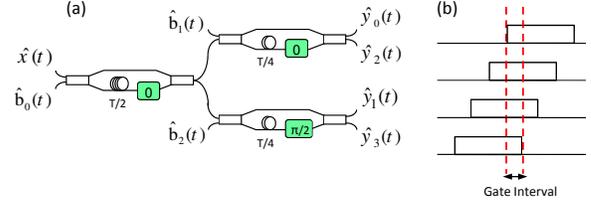}
\caption{(a) The passive ODFT circuit for $N=4$. The circuit consists of three MZIs with corresponding delays and phase shifts. (b) Shifted replicas of the input OFDM signal for $N=4$. The shift values for $N=4$ are $\{0,T/4,T/2,3T/4\}$. The time slot, in which all these copies overlap, is extracted by the time gating operation. \label{Fig:ODFT}} 
\end{figure}

Different methods can be used to realize the OFDM decoding, as required by (\ref{scheme1_DFT}), in the optical domain. In Scheme I, we assume that the ODFT is implemented by a passive structure consisted of $N-1$ MZIs \cite{hillerkuss2010simple}. Figure~\ref{Fig:ODFT}(a) shows an ODFT circuit for $N=4$. This structure imitates the efficient method of realizing DFT, known as fast Fourier transform (FFT), by means of delays, couplers, and phase shifters. Each output port of the ODFT circuit is a weighted sum of shifted replicas of the input as required by (\ref{scheme1_DFT}). It will then provide us with a real-time DFT operation once all shifted replicas of the input overlap, as shown in Fig.~\ref{Fig:ODFT}(b). That would require a time gating operation \cite{hillerkuss2010simple, hillerkuss201126}, which can be implemented by electro-absorption modulators (EAM), or simply by time-gating the single-photon detectors used in the QKD decoders. { Time misalignment can then be a major source of error in such a scheme.} The quantum operation of the ODFT circuit is discussed in more detail in Sec.~\ref{Sec:quantum}. 

\subsection{Scheme II}

\begin{figure}[!t]
\centering
\includegraphics[width=\linewidth]{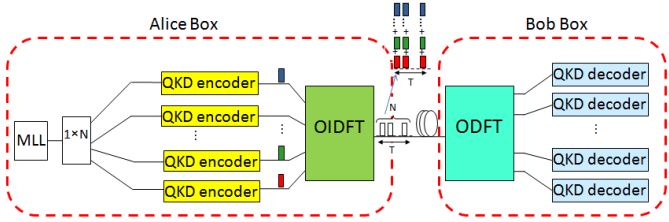}
\caption{OFDM-QKD using OIDFT circuit. A train of short pulses generated by an MLL is split into $N$ paths. The OFDM symbol is generated by multiplexing the output pulses of the QKD encoders by the OIDFT circuit. The OFDM symbol consists of a series of pulses, each a superposition of pulses from different inputs. At the receiver, an ODFT circuit demultiplexes the subcarriers. \label{Fig:scheme2}}
\end{figure}

Figure~\ref{Fig:scheme2} shows an alternative setup for the OFDM-QKD system. Here, the output of a pulsed laser source, e.g., a mode-locked laser (MLL), is split into several paths by an optical splitter. The pulses should be short enough to cover the spectrum of all the subcarriers in the OFDM symbol. Here, we assume that the pulse width is slightly lower than $T_c$. Similar to Scheme I, each short pulse, after splitting, is fed into QKD encoders to produce successive pulses $r$ and $s$. Each of these pulses will then go through an OIDFT circuit generating $N$ short pulses within an OFDM symbol duration $T$. The delay in the MZI of Fig.~\ref{Fig:phase_enc} is assumed to be greater than $T$.

The required OIDFT can be implemented by a structure similar to the ODFT. For instance, the circuit in Fig.~\ref{Fig:ODFT}(a), for the special case of $N=4$, can be employed for OIDFT as well. In the case of OIDFT, the input pulses (denoted by $y$ components) enter from the right hand side of Fig.~\ref{Fig:ODFT}(a) and the output will be the signal labeled by $\hat x(t)$. Assuming that the $y$ pulses are synchronous, in each OFDM symbol, the output signal $x$ consists of four pulses apart by multiples of $T_c$ within a $T$-long frame. Each of the latter pulses are a combination of all input pulses, as shown in Fig.~\ref{Fig:scheme2}. 

More generally, in the classical case, the generated OFDM amplitude at any carrier frequency $\omega$ is given by
\begin{equation}
x(t)=\sum_{k=0}^{N-1}{\sum_{l=0}^{N-1}{A_{k}p(t-l T_c)e^{j2\pi kl/N}}},
\end{equation}
where $p(t)$ represents the shape of the initial laser pulse and $A_k$ is the complex amplitude of the $k^{th}$ subchannel. Note that in Fig.~\ref{Fig:scheme2}, subchannels are separated spatially at the input to OIDFT.
At the receiver, the ODFT operation in (\ref{scheme1_DFT}) shifts each of the pulses within an OFDM symbol and combines them together to generate
\begin{equation}
y_{m}(t)=\frac{1}{N^2}\sum_{n=0}^{N-1}{\sum_{k=0}^{N-1}{\sum_{l=0}^{N-1}{A_{k}p(t-(n+l)T_c)e^{j2\pi (kl+nm)/N}}}},
\end{equation}
for $m\in\{ 0,1,...,N-1\}$. In a real-time implementation, only at $n+l=N-1$, all relevant input pulses are added together at which $y_m(t)$ reduces to
\begin{equation}
z_{m}(t)=\sum_{k=0}^{N-1}{\frac{A_{k}}{N}p(t-{(N-1)T_c})e^{j2\pi k(N-1)/N}(\frac{1}{N}\sum_{n=0}^{N-1}{e^{j2\pi n(m-k)/N}})}.
\end{equation}
Here again, the term in the brackets is zero for $k\neq m$, which implies that, up to a known overall phase factor, the original information in $A_m$ can be recovered at the $m$th output port of the ODFT circuit of Fig.~\ref{Fig:scheme2}.

\begin{figure}[!t]
\centering
\includegraphics[width=\linewidth]{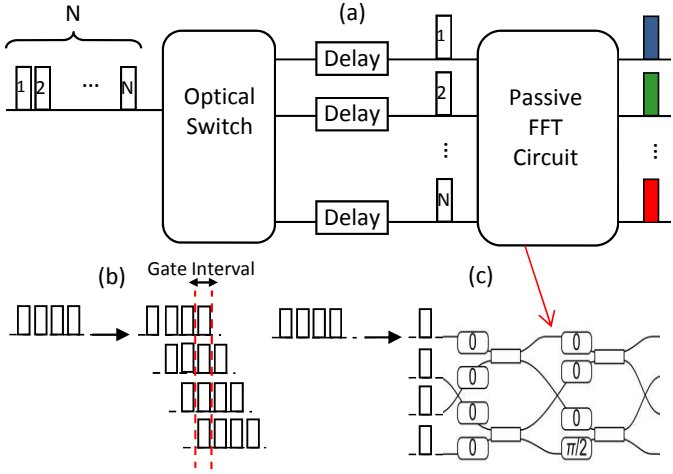}
\caption{(a) Proposed {\em active} ODFT circuit. By employing an optical switch instead of a power splitter, the loss of time gating is eliminated. (b) Passive approach to serial-to-parallel conversion for $N=4$. Some pulses are generated and then discarded during the time gating process. (c) Active approach to serial-to-parallel conversion for $N=4$ followed by the FFT circuit \cite{hillerkuss2010simple}. \label{Fig:activeODFT}} 
\end{figure}

For the receiver of Scheme II, we have two options. We can either use the passive OFDM decoder used in Scheme I, or, alternatively, the active structure shown in Fig.~\ref{Fig:activeODFT}(a). The main advantage of the latter is to remove the inherent loss in the passive OFDM decoder. To better explain the loss effect in the passive decoder, consider a sequence of $N$ pulses at the input $x(t)$ of Fig.~\ref{Fig:ODFT}(a), and let us look at the output signals. In this case, each input pulse has four paths to take, with different delays and phase shifts, to reach to the output ports of Fig.~\ref{Fig:ODFT}(a). In other words, for each input pulse, there will be four output pulses at each of the decoder's four output ports; see Fig.~\ref{Fig:activeODFT}(b). Only one out of these four output pulses has the right amount of delay and phase shift to be used for our ODFT operation, and that is why time gating is required. The inevitable drawback of this approach is that the other three pulses, and the power therein, will remain unused and that will contribute to a maximum total efficiency of $1/N$ for a passive decoder as in Fig.~\ref{Fig:ODFT}(a). To overcome this drawback, our proposed OFDM decoder in Fig.~\ref{Fig:activeODFT}(a) employs an optical switch along with proper delays, instead of a passive circuit, to perform the serial to parallel conversion. As shown in Fig.~\ref{Fig:activeODFT}(c), this way there will be no extra pulses to be discarded, and the ODFT process can be implemented by a passive $N$-by-$N$ circuit, of a star topology but with phase shifters along each internal path, with no fundamental overall loss \cite{Marhic:87, hillerkuss2010simple}. 

The { above} feature of the active decoder in Fig.~\ref{Fig:activeODFT}(a) makes it a better choice for high-rate QKD links, as we will see in the following sections. The passive decoder schemes can still be used for the sake of sharing the channel resources between multiple service providers. They do not, however, offer any total-rate advantage as compared to a single-carrier system. Note that the OFDM decoder of Fig.~\ref{Fig:activeODFT}(a) is mostly compatible with Scheme II, due to its discrete nature, as compared to Scheme I. { Nevertheless, one should take note of possible challenges of using the active decoder with Scheme II. While, by using a common pulsed source, Scheme II is more immune to phase noise or frequency offset problems, time misalignment is still a key concern. That is why, in the following sections, we study the impact of such timing errors in our OFDM-QKD setups. Secondly, while the active decoder removes the loss associated with time gating, its optical switch will introduce some additional insertion loss. The latter, within our practical regime of interest, is shown to be less than 2~dB for ultrafast optical switches and will be accounted for in our numerical analysis \cite{bogoni2012photonic}. Optical switches may also have nonzero extinction ratios, because of which some power leaks to other undesired output ports. This is a minor problem for the decoder of Fig.~\ref{Fig:activeODFT}(a), because the input pulses to the switch are non-overlapping in time. By using time gating and proper delay lines, the leaked power to other ports should not appear in the same time slot that time gating is taking place, hence has negligible effect on system performance.}

\section{Quantum Analysis} 
\label{Sec:quantum}
In this section, we analyze the OFDM-QKD systems proposed in Sec.~\ref{Sec:Description} from a quantum mechanical perspective.  We choose the Heisenberg picture for our analysis. In two steps, we first concentrate on the operation of the system corresponding to each of the two pulses $r$ and $s$ in Fig.~\ref{Fig:phase_enc}, and then we combine the results to find the output operators in QKD decoding modules. 

\subsection{Scheme I}  
In the Heisenberg picture, the output operator of the Alice box in Fig.~\ref{Fig:scheme1} can be expressed as 
\begin{equation}
\hat{x} (t)=\sum_{k=0}^{N-1} {\hat{a}_k e^{j\omega _{k} t}},\;\;\;\; 0 < t < T,
\label{InputSchemeI}
\end{equation}
where $\hat{a}_k$ is the annihilation operator corresponding to the mode representing the $k^{th}$ subcarrier. For the rest of this section, we neglect the path loss effect, which will be considered when we calculate the secret key generation rate. We then focus on the receiver setup assuming that at its input the signal $\hat x (t)$ is received. 

For simplicity, let us first consider the special case of $N=4$. As shown in Fig.~\ref{Fig:ODFT}, the ODFT circuit, in this case, is implemented by three MZIs. The operators $\hat{b}_{0} (t)=\sum_{k=0}^{N-1} {\hat{b}_{0k} e^{j\omega _{k} t}}$, $\hat{b}_{1} (t)=\sum_{k=0}^{N-1} {\hat{b}_{1k} e^{j\omega _{k} t}}$ and $\hat{b}_{2} (t)=\sum_{k=0}^{N-1} {\hat{b}_{2k} e^{j\omega _{k} t}}$ represent the vacuum fluctuations of the unused ports of the MZIs' beam splitters corresponding to all existing frequency modes of the system. For a center frequency, $\omega$, the transformation matrices of the three MZIs in Fig.~\ref{Fig:ODFT} are given by
\begin{equation}
B_{\omega ,1}=\frac{1}{2}
\left (
\begin{array}{cc}
1& j\\j &1
\end{array}
\right )
\left (
\begin{array}{cc}
1& 0\\0 &e^{-j(\omega \frac{T}{2})}
\end{array}
\right )
\left (
\begin{array}{cc}
1& j\\j &1
\end{array}
\right ),
\end{equation}
for the MZI on the left,
\begin{equation}
B_{\omega ,2}=\frac{1}{2}
\left (
\begin{array}{cc}
1& j\\j& 1
\end{array}
\right )
\left (
\begin{array}{cc}
1& 0\\0 &e^{-j(\omega \frac{T}{4})}
\end{array}
\right )
\left (
\begin{array}{cc}
1& j\\j& 1
\end{array}
\right ),
\end{equation}
for the one on top right, and
\begin{equation}
B_{\omega ,3}=\frac{1}{2}
\left (
\begin{array}{cc}
1& j\\j& 1
\end{array}
\right )
\left (
\begin{array}{cc}
1& 0\\0& je^{-j(\omega \frac{T}{4})}
\end{array}
\right )
\left (
\begin{array}{cc}
1& j\\j& 1
\end{array}
\right ),
\end{equation}
for the one on bottom right of Fig.~\ref{Fig:ODFT}(a). Applying the above transformations to mode $k$, we obtain
\begin{equation}
\left (
\begin{array}{c}
\hat{a}^{\prime}_{k} (t) \\
\hat{b}^{\prime}_{k} (t) 
\end{array}
\right )
=B_{\omega_{k} ,1}
\left (
\begin{array}{c}
\hat{a}_{k} (t) \\
\hat{b}_{0k} (t) 
\end{array}
\right ),
\end{equation}
\begin{equation}
\left (
\begin{array}{c}
\hat{y}_{0,k} (t) \\
\hat{y}_{1,k} (t) 
\end{array}
\right )
=B_{\omega_{k} ,2}
\left (
\begin{array}{c}
\hat{a}^{\prime}_{k} (t)\\
\hat{b}_{1k} (t) 
\end{array}
\right ),
\end{equation}
\begin{equation}
\left (
\begin{array}{c}
\hat{y}_{2,k} (t) \\
\hat{y}_{3,k} (t) 
\end{array}
\right )
=B_{\omega_{k} ,3}
\left (
\begin{array}{c}
\hat{b}^{\prime}_{k} (t)\\
\hat{b}_{2k} (t) 
\end{array}
\right ).
\end{equation}
The output operator for output port $m$ in Fig.~\ref{Fig:ODFT} is then given by
\begin{equation}
\hat{y}_{m} (t)=\sum_{k=0}^{3}{\hat{y}_{m,k} (t)}, \;\;\; m=0,1,2,3 .
\end{equation}
Note that the above operations are all linear. Based on the superposition principle, we can split each output $\hat{y}_m (t) $ to two parts. The first part is the output obtained by neglecting the vaccum operators, and the other part is a linear combination of all  vacuum operators. More generally, it can be concluded that the output of such an ODFT circuit, neglecting the vacuum operators, can be expressed as a function of $\hat{x}(t)$, as follows:
\begin{equation}
\hat{X}_m (t)=\frac{1}{N}\sum_{n=0}^{N-1}{\hat{x}(t- n T_c) e^{j2\pi nm/N}}, \quad\mbox{$m= 0,\ldots, N-1$},
\label{ODFT}
\end{equation}
which is similar in form to (\ref{scheme1_DFT}) for the classical case. Substituting (\ref{InputSchemeI}) into (\ref{ODFT}) and applying the orthogonality condition, $\Delta f=\frac{1}{T}$, we obtain
\begin{equation}
\hat{X}_m (t)= \sum_{k=0}^{N-1}{\hat{a}_{k} e^{j2\pi(f_0 +k \Delta f)t} (\frac{1}{N}\sum_{n=0}^{N-1}{e^{j2\pi n (m-k)/N}})}.
\end{equation}
Note that the term in the brackets is nonzero only if $k=m$. We then obtain
\begin{equation}
\hat{y}_{m} (t)=A\hat{a}_{m} (t)+\sum_{i=0}^{N-2}{\sum_{k=0}^{N-1}{\beta_{ik}\hat{b}_{ik}(t)}}, \quad\mbox{$m= 0,\ldots, N-1$},
\end{equation}
where $A$ is either $1$ or $j$ and $\beta_{ik}$'s are constant coefficients. The operator $\hat{a}_{m} (t)$ is the evolved version of $\hat{a}_{m}$ and is given by $\hat{a}_{m} (t) = \hat{a}_{m} e^{j\omega _ {m}t}$, and similarly for the vacuum operators in the above equation. As explained in Sec.~\ref{Sec:Description}, the orthogonality is only met in a region of width $T/N$, where all of the shifted copies of the OFDM signal overlap. The signal corresponding to this overlapping time slot will eventually be detected by the photodetectors in the receiver module. 

With the phase encoding QKD protocol, the two successive pulses $r$ and $s$ for channel $m$, represented by $\hat{a}_{m}^{(r)}$ and $\hat{a}_{m}^{(s)}$, respectively, will be recombined at the receiver's MZI in Fig.~\ref{Fig:phase_enc}. The output operator corresponding to the recombined pulse $d$ in Fig.~\ref{Fig:phase_enc} for the $m^{th}$ output is then given by
\begin{eqnarray}
\hat{d}_{m}(t)=\frac{j}{2}  (e^{j\phi _{B}^{(m)}}\hat{a}_{m}^{(r)}(t)+\hat{a}_{m}^{(s)}(t))+\sum_{i=0}^{N-2}{\sum_{k=0}^{N-1}{\beta_{ik}^{(r)}\hat{b}_{ik}^{(r)}(t)}}+\nonumber \\
\sum_{i=0}^{N-2}{\sum_{k=0}^{N-1}{\beta_{ik}^{(s)}\hat{b}_{ik}^{(s)}(t)}},\;\;\;\; \;\;\; T - T_c <t<T.
\label{d_schemeI}
\end{eqnarray} 
\subsection{Scheme II}
To start our analysis in this scheme, we denote the annihilation operator corresponding to the spatial mode at the output of the $k^{th}$ QKD encoder by $\hat{a}_{k}$. We assume that an OIDFT circuit similar to the one depicted in Fig.~\ref{Fig:ODFT}(a) for $N=4$, yet in the reverse direction, is used at the transmitter. Then, we can obtain the output operator of the Alice box by applying the transformation matrix of each MZI. It can be concluded that the output operator of the Alice box is given by
\begin{equation}
\hat{x}(t)=\frac{1}{N}\sum_{l=0}^{N-1}{\hat{c}_{l}p(t-l T_c)},
\label{InputSchemeII}
\end{equation}
where $\hat{c}_{l}=\sum_{k=0}^{N-1}{\hat{a}_{k}e^{j2\pi kl/N}}$ is the $l^{th}$ temporal mode at the output of the OIDFT circuit. Note that the coefficient $1/N$ in (\ref{InputSchemeII}) is not necessarily a source of loss, so long as the average number of photons per pulse at the output of the transmitter meets the requirements of the decoy-state protcol. That is, we can compensate for the internal loss at the transmitter by tuning the intensity of the input light. In our following analysis, this factor $1/N$ has been neglected. With a passive OFDM decoder similar to that of Scheme I at the receiver, the analysis presented in the previous subsection can be useful here as well, except that here we deal with temporal modes. Substituting (\ref{InputSchemeII}) in (\ref{ODFT}) and simplifying the equations at $n+l=N-1$, we can express each output of the ODFT circuit as
\begin{eqnarray}
\hat{y}_{m}(t)=[\hat{a}_{m}e^{j2\pi m(N-1)/N}+\sum_{i=0}^{N-2}{\sum_{k=0}^{N-1}{\beta_{ik}\hat{b}_{ik}}}]\nonumber\\
\times p(t-T+T_c), \quad\mbox{$m= 0,\ldots, N-1$}.
\end{eqnarray}
The coefficient $e^{j2\pi m(N-1)/N}$ is a constant phase term that can be absorbed in $\hat{a}_m$ in the above equation. 

Finally, the output operator obtained by the recombination of the pulses $r$ and $s$ by means of the receiver's MZI, $\hat{d}_{m}(t)$, is given by
\begin{eqnarray}
\hat{d}_{m}(t)=[\frac{j}{2}  (e^{j\phi _{B}^{(m)}}\hat{a}_{m}^{(r)}+\hat{a}_{m}^{(s)})+\sum_{i=0}^{N-2}{\sum_{k=0}^{N-1}{\beta_{ik}^{(r)}\hat{b}_{ik}^{(r)}}}+\nonumber \\
\sum_{i=0}^{N-2}{\sum_{k=0}^{N-1}{\beta_{ik}^{(s)}\hat{b}_{ik}^{(s)}}}]\times p(t-T+ T_c).
\label{d_schemeII}
\end{eqnarray}   

Another option for the receiver in this scheme is the structure we proposed in Fig.~\ref{Fig:activeODFT}(a). For this active decoder, (\ref{d_schemeII}) is multipled by an additional factor $\sqrt{N}$. Furthermore, no vacuum components would appear because the beam splitters in the FFT circuit do not have any unused ports \cite{hillerkuss2010simple}.

\section{Key Rate Analysis}
\label{Sec:keyrate}

This section presents an analysis of the secret key generation rate for the proposed OFDM-QKD schemes. We assume that the efficient decoy-state BB84 protocol is employed in the QKD setup \cite{lo2005efficient, panayi2014memory}. 
The average number of photons per QKD channel is given by $\mu$, for the main signal state, and it is calculated at the output of the Alice box. The secret key generation rate per transmitted pulse, in the limit of an infinitely long key, is lower bounded by $\max [0,P(\mathrm{Y}_0)]$, where 
\begin{equation}
P(\mathrm{Y}_0)=Q_{1} (1-h(e_1))-f Q_{\mu}h(E_{\mu}),
\label{Y0}
\end{equation}
and $h(p)=-p{\log}_{2}p-(1-p){\log}_{2}(1-p)$ is the binary entropy function with $f$ being the error correction inefficiency. The overall gain, the QBER, the gain of a single photon state and the error rate of a single photon state are, respectively, given by \cite{Razavi_MulipleAccessQKD}
\begin{eqnarray}
Q_{\mu}=1-(1-\mathrm{Y}_{0})e^{-\eta \mu},
E_{\mu}=(\mathrm{Y}_{0}/2+e_{d}(1-e^{-\eta\mu})) /Q_{\mu},  \nonumber\\
Q_{1}=\mathrm{Y}_{1}\mu e^{-\mu},
 e_{1}=(\mathrm{Y}_{0}/2+e_{d}\eta)/ \mathrm{Y}_{1}.\;\;\;\;\;\;\;\;\;\;
\label{gain-error}
\end{eqnarray}
Equations (\ref{Y0})-(\ref{gain-error}) provide an estimate to the generated key rate when infinitely many decoy states are in use and no eavesdropper is present. In the above equations, $\mathrm{Y}_{1}=(1-\eta)\mathrm{Y}_{0}+\eta$ is the yield of a single photon state and $\mathrm{Y}_{0}$ is the probability of any detector clicks without having any transmitted photons from the corresponding QKD encoder. Furthermore, $e_d$ represents the probability of phase stability errors, { between $r$ and $s$ pulses}, and the total transmissivity of the link is given by
\begin{equation}
\eta=\eta_{g} \eta_d \eta_{\rm ins} 10^{-\alpha L/10},
\end{equation}
where $\eta_d$ is the quantum efficiency of the photodetectors, $\alpha$ is the channel loss factor in dB per unit of length and $\eta_{\rm ins}$ represents any additional insertion loss in the link. Here, $\eta_g$ represents the additional loss due to the OFDM decoding scheme. For instance, in Scheme I, with gate interval of $T_c$, the parameter $\eta_{g}$ equals $1/N$ in the ideal case. For Scheme II, and the active decoder of Fig.~\ref{Fig:activeODFT}(a), $\eta_g$ is ideally one. 

We calculate the parameters in (\ref{gain-error}) by finding the probabilities of interest once the QKD measurements are done. For instance, the measurement operator for the representative photodetector in Fig.~\ref{Fig:phase_enc} is given by
\begin{equation}
\hat{M}=\int_{ { gate}\;{inteval}}{\hat{d}_{m}^{\dag} (t)\hat{d}_{m} (t)dt}, 
\end{equation}
from which one can obtain key rate parameters. In the analysis of the key generation rate, the terms containing vacuum states will not contribute to the key rate parameters, and that simplifies the calculations. 

In order to analyze the secret key generation rate of the proposed schemes in more detail, one should consider the influence of imperfections in the system, which may degrade system performance. As explained before, we specifically consider time misalignment issues, which are known to be critical in all-optical OFDM systems. In OFDM, the time alignment of the optical subchannels is critical, due to its effect on their orthogonality. Furthermore, the time gating at the receiver should be synchronized with the transmitted pulses to extract the correct time slot. In our OFDM-QKD setups, nonidentical QKD encoders or some errors in time-gating synchronization may introduce time misalignment.


In our work, we have found the key generation rate of our proposed OFDM-QKD schemes in the presence of time misalignment issues. It turns out that they cause two problems. First, they generate some inter-channel crosstalk, denoted by $p_{\rm xtalk}$, which adds to the background noise, and, second, they slightly reduce the transmissivity factor $\eta_g$. One can reduce the crosstalk noise by reducing the width of the gate interval, but, by doing so, $\eta_g$ would further be reduced, as we have to leave out some of the desired signal components as well. That would imply the existence of an optimal gate width at which the total secret key rate $R_{\rm OFDM}$ is maximized, where 
\begin{equation}
{R}_{\mathrm{OFDM}}=\max[NP(\mathrm{Y}_{\mathrm{OFDM}})/T_{s},0],
\label{R_OFDM}
\end{equation}
and
\begin{equation}
\mathrm{Y}_{\mathrm{OFDM}}=1-(1-(p_{\rm dc}+p_{\rm xtalk}))^2.
\label{Y_OFDM}
\end{equation}
Here, $T_s$ represents the repetition period of the QKD protocol { and $p_{\rm dc}=\gamma_{\rm dc}{T_g}$, where $\gamma_{\rm dc}$ and $T_g$ are the photodetectors' dark count rate and the gate interval, respectively.} In Appendix~A, we have derived all the required terms for calculating the above key rate as a function of the average time misalignment $\mathrm{E}\{|\tau _{k}|\}$, where $\tau _{k}$'s are i.i.d random variables representing the time misalignment of the $k$th channel with respect to the gate interval, and $\delta\triangleq (T_c -T_p)/2$, where $T_p$ is the pulse width in Scheme II. In the following, we present some of our numerical findings for a selected set of parameter values.

\section{Numerical Results}

\label{Sec:num}
\begin{table}[pbt]
\caption{Nominal values for system parameters}
\centering 
\begin{tabular}{|c |c |} 
\hline
Parameter & Value\\
\hline
Average number of photons per signal pulse & 0.48\\
Quantum Efficiency & 0.3\\
Total insertion and path loss, $\eta_{\rm ins}10^{-\alpha L/10}$& 10 dB\\
Receiver dark count rate, ${\gamma} _{dc}$& 1E-7 ${\rm ns}^{-1}$\\
Error correction inefficiency, $f$& 1.22\\
Phase stability error, $e_d$& {0.005}\\
Laser pulse repetition interval, $T_s$ & 210 ${\rm ps}$\\
OFDM symbol duration, $T$ & 100 ${\rm ps}$\\
Number of subcarriers, $N$ & 4, 8, 16\\
\hline
\end{tabular}
\end{table}

{In this section, we investigate the performance of the proposed OFDM-QKD schemes by considering the following cases: Scheme I and Scheme II with passive OFDM decoders, and Scheme II with the active OFDM decoder of Fig.~\ref{Fig:activeODFT}(a). In order to evaluate the effect of time misalignment on the performance of each case, each subcarrier is assumed to have a time misalignment with uniform distribution, $\tau_k \sim \mathrm{U} (-a,a)$, where $a< T_c$ is an arbitrary constant. { We then find the optimal gate width that maximizes the key rate.} The nominal values used for the system parameters are listed in Table I. These parameters are chosen in accordance to practical considerations. With the chosen OFDM symbol duration, the subchannel frequency separation, $\Delta f$, has to be 10 GHz, which has been used in several all-optical OFDM experiments \cite{wang2011optical}, \cite{lowery2011all}. The pulse width, $T_p$, in Scheme II should be less than $T_c=T/N$. Here, we assume that the ratio $\delta / T_p$ in this scheme is equal to { 0.04}.

\begin{figure}[!t]
\centering
\includegraphics[width=\linewidth]{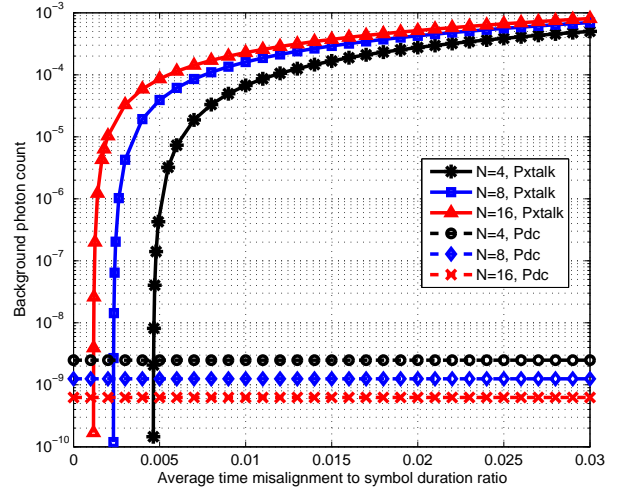}
\caption{{ Background photon count probability components $p_{\rm dc}$ and $p_{\rm xtalk}$ versus $\mathrm{E}\{|\tau_{k}|\}/T$ for Scheme II with active OFDM decoder, for different values of $N$. } \label{Pdc_xtalk}}
\end{figure}

{ In order to see the importance of the time misalignment issue, we first look at its induced cross talk contribution as compared to the dark count component. Figure~\ref{Pdc_xtalk} compares the two elements of the background noise, i.e., $p_{\rm dc}$ and $p_{\rm xtalk}$, versus a normalized measure of misalignment, $\mathrm{E}\{|\tau_{k}|\}/T$, in the special case of Scheme II with active decoders. A similar overall behavior is observed for other schemes as well. It is clear that while the cross talk is negligible for low values of time misalignment, it becomes the major source of noise in our OFDM-QKD setups. We next consider the effect of time misalignment on each of the proposed setups.}

\begin{figure}[!t]
\centering
\includegraphics[width=\linewidth]{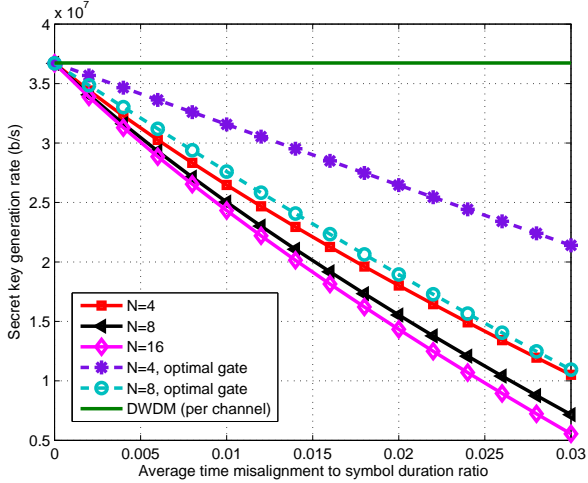}
\caption{{ Secret key generation rate versus $\mathrm{E}\{|\tau_{k}|\}/T$ for Scheme I and its optimal gate version, for different values of $N$.} \label{scheme1}}
\end{figure}

Figure~\ref{scheme1} shows the effect of time misalignment on the total secret key generation rate, $R_{\rm OFDM}$, of Scheme I with and without optimal time gating. In the latter case, the gate width is a constant $T_c$. It can be seen that by optimizing the gate interval the secret key rate significantly improves. It will not, however, surpass the performance of a single-carrier link, { shown on top of Fig.~\ref{scheme1},} run at the same clock rate as the OFDM system. The main reason for this is the loss factor $N$ due to the time gating, which results in a reduced key rate per carrier. Moreover, Fig.~\ref{scheme1} shows that a system with a larger number of subcarriers, $N$, is more susceptible to time misalignment errors. This has to do with the interplay between $p_{\rm xtalk}$ and $\eta_g$, where the latter turns out to be the dominant factor. In short, no rate advantage is obtained by Scheme I. It, nevertheless, can be used as a multiplexing tool for sharing the infrastructure between multiple service providers. 

\begin{figure}[!t]
\centering
\includegraphics[width=\linewidth]{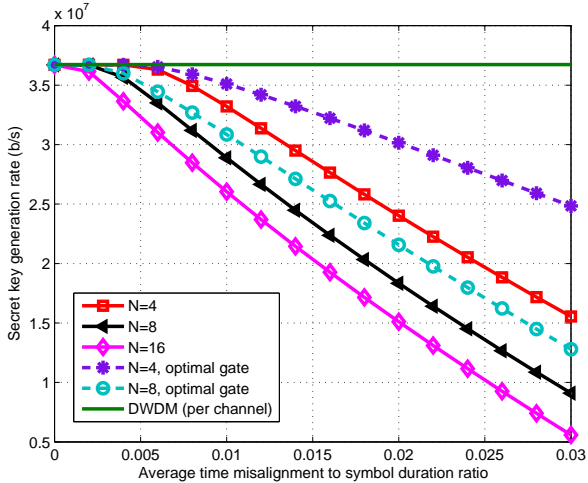}
\caption{{ Secret key generation rate versus $\mathrm{E}\{|\tau_{k}|\}/T$ for Scheme II with passive OFDM decoder and its optimal gate version, for different values of $N$. The parameter $\delta/T_p$ is chosen to be 0.04.}\label{scheme2_p} }
\end{figure} 

Next, Fig.~\ref{scheme2_p} shows the total secret key rate of Scheme II with passive OFDM decoder versus $\mathrm{E}\{|\tau_{k}|\}/T$. Here again, applying the optimal gate interval results in an enhancement in the secret key rate. Yet, no improvement, as compared to the DWDM-QKD system, is observed in the overall key rate by increasing $N$, which is mainly due to the inherent loss in the passive structure of the OFDM decoder.

The secret key generation rate of Scheme II with the proposed active OFDM decoder is depicted in Fig.~\ref{scheme2_a}. It can be seen that by multiplexing more subchannels in the system the secret key rate increases. This increase is initially linear with the number of subchannels, but once the time misalignment kicks in the key rate also correspondingly drops. { Nevertheless, it always stays above that of the single DWDM-QKD channel depicted in the bottom of the figure. In fairness to the DWDM system, we have accounted for 2~dB of additional insertion loss for the optical switch in the OFDM-QKD system. The DWDM curve uses 100-ps-long pulses.} Under these conditions, for $N=16$, and at a normalized average time misalignment of 0.02, we are doing more than four times better than the single carrier link. That would demonstrate the prospect of using OFDM techniques at the core of QKD networks. 

\begin{figure}[!h]
\centering
\includegraphics[width=\linewidth]{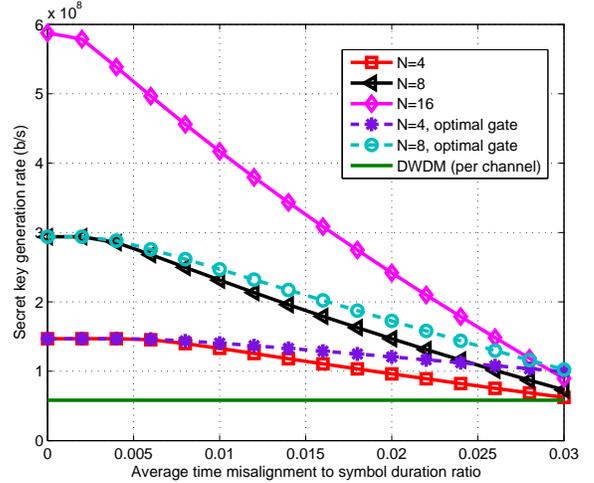}
\caption{{ Secret key generation rate versus $\mathrm{E}\{|\tau_{k}|\}/T$ for Scheme II with active OFDM decoder and its optimal gate version, for different values of $N$. The parameter $\delta/T_p$ is chosen to be 0.04.} \label{scheme2_a}}
\end{figure} 

{ In addition to the total key rate, we also look at the spectral efficiency of each scheme, $S$, defined by the ratio of the secret key generation rate and the allocated bandwidth. In the case of a DWDM link with channel spacing of 50~GHz, in Fig.~\ref{scheme2_a}, $S = 0.11 \%$. For the OFDM-QKD systems, assuming that the allocated bandwidth is given by $N/T$, $S$ has a peak value of 0.36\%, which is three times higher than that of the DWDM system. Once time misalignment kicks in, the OFDM-QKD systems with lower values of $N$ are favored as they are less susceptible to such errors. Overall, it can be seen that by multiplexing 4--8 OFDM subcarriers, one can outperform DWDM-QKD systems both in terms of the total key rate and the spectral efficiency in practical regimes of operation.}
  
\section{Conclusions}
We proposed a spectrally efficient approach to multiplexing QKD channels, namely, OFDM-QKD. 
Based on the principles of all-optical OFDM in classical communications, several OFDM-QKD schemes were considered. These schemes were analyzed in detail, in terms of their secret key generation rate, considering time-alignment imperfections, which are critical in all-optical OFDM systems. It was shown that such time misalignment issues would introduce a crosstalk noise with a degrading effect on the key rate, similar to that of background noise. We showed that by reducing the gate interval to an optimal value this problem could be alleviated to a large extent. Most importantly, we showed that the existing passive structures for the OFDM decoder would provide no gain in their multiplexing, in terms of the total achievable key rate. We proposed an active OFDM decoder, which, by using an optical switch, followed by proper delays and a passive FFT circuit, could eliminate the inherent loss in passive decoders. We remark that, in the case of active decoders, ultrafast switches with a transition time on the order of picoseconds may be required. This is due to the short time separation of pulses within an OFDM symbol \cite{bogoni2012photonic,chou2004compact}. { This may add to the cost and complexity of the system. Nevertheless, we showed that, using our proposed active decoders, we could outperform the alternative DWDM-QKD systems in terms of the total key rate and spectral efficiency.} This implies that OFDM-QKD can provide a high-rate spectrally efficient method of key exchange at the core of trusted-node QKD networks.  

\appendices
\section{OFDM-QKD with misalignment errors}
In this appendix, we analyze the operation of our proposed OFDM-QKD setups in the presence of time misalignment.
\subsection{Scheme I}
We start our analysis by assuming that each subcarrier has a time misalignment $\tau _{k}$, $0<|\tau _{k}|< T_c$, with respect to the time gating interval. Without loss of generality, we assume that $0\leq \tau _{k}<  T_c$ (both cases of $\tau _{k}>0$ and $\tau _{k}<0$ have the same effect). 
Figure~\ref{misalignment}(a) shows the shifted copies of the $k^{th}$ subcarrier pulse in the presence of time misalignment $\tau_k$. As can be seen in the figure, the shifted copies does not overlap completely in the gate interval, which leads to different summation results in two distinct time intervals, as follows:\\
\begin{eqnarray}
t\in (T-T_{c},T-T_{c}+\tau _{k}) \Rightarrow \hat{X}_m (t)=\left\{ 
\begin{array}{cc}
\frac{1}{N}\hat{a}_k (t)& k\neq m\\
\frac{N-1}{N}\hat{a}_{m} (t) & k= m 
\end{array}
\right.\\
t\in (T-T_{c}+\tau _{k}, T)  \Rightarrow \hat{X}_m (t)=\left\{ 
\begin{array}{cc}
0& k\neq m\\
\hat{a}_{m} (t)& k= m 
\end{array}
\right. .
\end{eqnarray} 
As a consequence, equation (\ref{d_schemeI}) is modified to
\begin{eqnarray}
\hat{d}_{m}(t)=\frac{j}{2}\{\frac{N-1}{N}(e^{j\phi _{B}^{(m)}}\hat{a}_{m}^{(r)}(t)+\hat{a}_{m}^{(s)}(t))\vert _{t\in (T-T_c,T-T_c+\tau _{m}) }+ \nonumber \\ (e^{j\phi _{B}^{(m)}}\hat{a}_{m}^{(r)}(t)+\hat{a}_{m}^{(s)}(t))\vert _{t\in (T-T_c+\tau _{m},T) }+ \nonumber\\
\frac{1}{N}\sum_{k\neq m}{(e^{j\phi _{B}^{(k)}}\hat{a}_{k}^{(r)}(t)+\hat{a}_{k}^{(s)}(t))\vert _{t\in (T-T_c,T-T_c+\tau _{k}) }}\},
\label{d_SchemeI_t}
\end{eqnarray}
with the third term representing the inter-subcarrier crosstalk on the $m^{th}$ subcarrier. Here, we neglected the vacuum operators due to their elimination once we apply the measurement operator. The background count due to this crosstalk may influence the performance of the system, as we will show in the following.

Defining $\hat{g}_{k}(t)\triangleq \frac{1}{N}(e^{j\phi _{B}^{(k)}}\hat{a}_{k}^{(r)}(t)+\hat{a}_{k}^{(s)}(t))$ with initial state $|\alpha\rangle _{r}|\alpha e^{j\phi _{A}^{(k)}}\rangle _{s}$, we can write
\begin{equation}
\langle\hat{g}_{k}^{\dag}\hat{g}_{k}\rangle=\frac{2 \mu}{N^2} (1+cos(\phi _{A}^{(k)}-\phi _{B}^{k}))\frac{\tau _{k}}{T}.
\label{g}
\end{equation}
Here, $\mu= {|\alpha|}^{2}$ and $\phi _{A}^{(k)}$ is the relative phase produced by the QKD encoder of the $k^{th}$ subcarrier. We then calculate the expected value of (\ref{g}) as a function of $(\phi _{A}^{(k)}-\phi _{B}^{(k)})$, which results in 
\begin{eqnarray}
\mathrm{E}_{(\phi _{A}^{(k)}-\phi _{B}^{k})}\{\langle\hat{g}_{k}^{\dag}\hat{g}_{k}\rangle \}=\frac{1}{2}\mathrm{E}_{(\phi _{A}^{(k)}-\phi _{B}^{(k)})}\{\langle\hat{g}_{k}^{\dag}\hat{g}_{k}\rangle  | basis=\{0,\pi \}\}+\nonumber \\
\frac{1}{2}\mathrm{E}_{(\phi _{A}^{(k)}-\phi _{B}^{(k)})}\{\langle\hat{g}_{k}^{\dag}\hat{g}_{k}\rangle  | basis=\{ \pi /2, 3\pi /2 \}\}=\frac{2\mu\tau _{k}}{N^2 T}.
\label{av_g}
\end{eqnarray}
Note that cross terms between any two different subcarriers also appear in the $\langle\hat{d}_{m}^{\dag}\hat{d}_{m}\rangle$. Here, we assume that the laser sources have independent phases. In this case, the phase difference corresponding to any cross term has uniform distribution on the interval $[-\pi,\pi]$. These terms are then eliminated due to their zero expected values. At this point, we generalize our result to include the case $-T_c\leq \tau _{k}< 0$. Hence, equation (\ref{av_g}) can be rewritten as 
\begin{equation}
\mathrm{E}_{(\phi _{A}^{(k)}-\phi _{B}^{k})}\{\langle\hat{g}_{k}^{\dag}\hat{g}_{k}\rangle \}=\frac{2\mu|\tau _{k}|}{N^2 T}.
\end{equation}
In the last step, the partial crosstalks due to each subcarrier are added to obtain the total crosstalk background count on the $m^{th}$ subcarrier, denoted by $p_{\rm xtalk}^{(m)}$, as follows: 
\begin{equation}
p_{\rm xtalk}^{(m)}=\eta^{\prime}\frac{2\mu }{N^2 T}\sum_{k\neq m}{\mathrm{E}\{|\tau _{k}|\}} ,
\end{equation}
where $\eta^{\prime}$ is the transmissivity of the link, excluding the loss of time gating. Under the assumption that $|\tau _{k}|$'s are i.i.d random variables, $p_{\rm xtalk}^{(m)}$ is independent of the subcarrier index. So, we can express the crosstalk background count per subcarrier as
\begin{equation}
p_{\rm xtalk}=\eta^{\prime}\frac{2\mu (N-1)\mathrm{E}\{|\tau _{k}|\}}{N^2 T}.
\end{equation} 
Such time misalignments change the loss factor in the time gating operation, $\eta_{g}$, due to the additional loss  occurring in the interval $(T-T_c,T-T_c+\tau _{m})$ in (\ref{d_SchemeI_t}). $\eta_{g}$ is then given by
 \begin{equation}
\eta_g=\frac{1}{N}-(\frac{{ \mathrm{E}\{|\tau _{k}|\}}}{T})(1-(\frac{N-1}{N})^2).
\end{equation}

{Now, let us reduce the gate interval by $b$ from each side to reduce the crosstalk effect. It can be concluded from Fig.~\ref{misalignment}(a) that $p_{\rm xtalk}=\mathrm{E}\{A\}$, where $A$ is obtained by
\begin{equation}
A=\left\{
\begin{array}{cc}
 \eta^{\prime}\frac{2\mu (N-1)}{N^2 T}(|\tau _{k}|-b)& |\tau _{k}|\geq b\\
0 & |\tau _{k}|\leq b
\end{array}
\right.
\end{equation}
and $p_{\rm xtalk}$ can be expressed as
\begin{equation}
p_{\rm xtalk}=\eta^{\prime}\frac{2\mu (N-1)}{N^2 T}p(|\tau _{k}|\geq b)(\mathrm {E}\{|\tau _{k}|||\tau _{k}|\geq b\}-b).
\end{equation}
Furtheremore, $\eta_g=\mathrm{E}\{B\}$, where
\begin{equation}
B=\left\{
\begin{array}{cc}
\frac{1}{N}-\frac{2b}{T}-\frac{1}{T}(1-(\frac{N-1}{N})^2)((|\tau _{k}|-b)& |\tau _{k}|\geq b\\
0 & 0\leq |\tau _{k}|\leq b\\
\end{array}
\right.
\end{equation}
and
\begin{equation}
\eta_g=\frac{1}{N}-\frac{2b}{T}-\frac{1}{T}(1-(\frac{N-1}{N})^2)p(|\tau _{k}|\geq b)(\mathrm {E}\{|\tau _{k}|||\tau _{k}|\geq b\}-b).
\end{equation}
}
\subsection{Scheme II}
As explained in Sec.~\ref{Sec:Description}, two receiver structures can be applied for this scheme: the passive OFDM decoder used in Scheme I, and the active OFDM decoder, which exploits an optical switch. In this subsection, we derive the crosstalk background count for both receiver structures. 
\begin{figure}[!t]
\centering
\includegraphics[width=\linewidth]{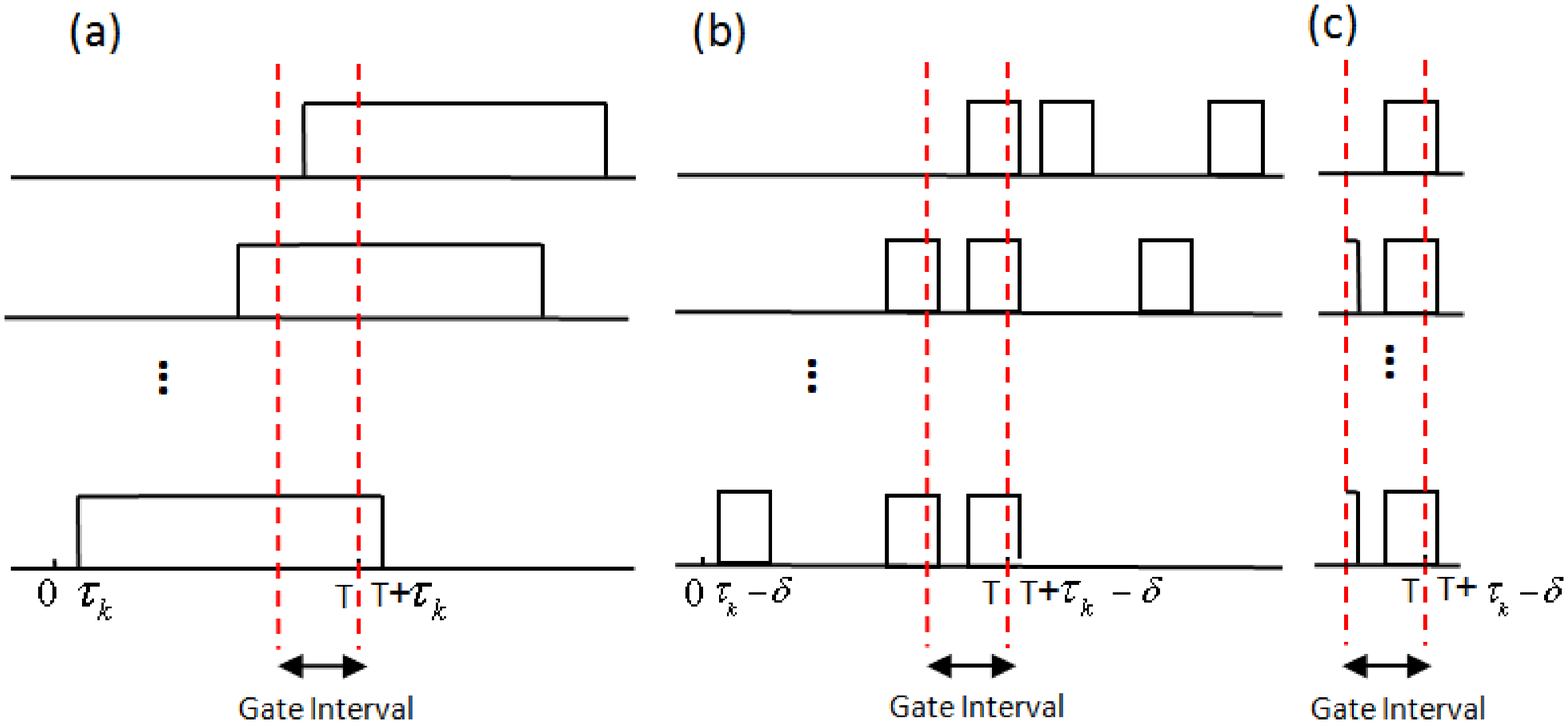}
\caption{Shifted copies of the signal corresponding to $k^{th}$ tributary in the presence of the time misalignment $\tau_{k}$, for (a) Scheme I, (b) Scheme II with passive OFDM decoder, (c) Scheme II with active OFDM decoder. \label{misalignment}} 
\end{figure}

First, we consider Scheme II with a passive OFDM decoder. We assume a time misalignment $\tau _{k}$, $0<|\tau _{k}|<T_c$, with respect to the time gating interval for each tributary. Figure~\ref{misalignment}(b) depicts the replicas of the pulse series corresponding to the $k^{th}$ tributary. We denote the width of each pulse by $T_p$. Following the same steps as in the previous subsection, we conclude that the output operator, $\hat{d}_{m}(t)$, for $0<\tau _{k}<\delta$ is given by
\begin{equation}
 \hat{d}_m (t) =\frac{j}{2}(e^{j\phi _{B}^{(m)}}\hat{a}_{m}^{(r)}(t)+\hat{a}_{m}^{(r)}(t)),
\label{d_SchemeII_t1}
\end{equation}
and for $\delta<\tau _{k}<T_c$ we have
\begin{eqnarray}
 \hat{d}_m (t)&=& \frac{j}{2}\{\frac{N-1}{N}(e^{j\phi _{B}^{(m)}}\hat{a}_{m}^{(r)}(t)+\hat{a}_{m}^{(s)}(t))\vert _{t\in (T-T_c,T-T_c+\tau _{m}-\delta)}\nonumber \\
&+&(e^{j\phi _{B}^{(m)}}\hat{a}_{m}^{(r)}(t)+\hat{a}_{m}^{(s)}(t))\vert _{t\in (T-T_c+\tau _{m}+\delta, T) }\nonumber \\
&+&\frac{1}{N}\sum_{k\neq m}{(e^{j\phi _{B}^{(m)}}\hat{a}_{k}^{(r)}(t)+\hat{a}_{k}^{(s)}(t))\vert _{t\in (T-T_c,T-T_c+\tau _{k}-\delta) }}\},
\label{d_SchemeII_t}
\end{eqnarray}
where $\delta=(T_c-T_{p})/2$. From this equation we can conclude that in the case of $\delta<|\tau _{k}|<T_c$, an inter-subcarrier crosstalk is introduced. We can then derive the background count of such crosstalk by applying the same strategy as in the previous subsection. The final result can be expressed as
\begin{equation}
p_{\rm xtalk}=\eta^{\prime}\frac{2\mu (N-1)}{N^3T_p}p(|\tau _{k}|\geq \delta)(\mathrm{E}\{|\tau _{k}|||\tau _{k}|>\delta \}-\delta).
\label{P_xtalk_II}
\end{equation} 
Furthermore, the degrading effect of time misalignment on $\eta_g$ modifies this parameter to 
 \begin{equation}
\eta_g=\frac{1}{N}-(\frac{1}{NT_p})(1-(\frac{N-1}{N})^2)p(|\tau _{k}|\geq \delta)(\mathrm{E}\{|\tau _{k}|||\tau _{k}|>\delta \}-\delta).
\label{eta_g_s_2_p}
\end{equation}

{
Now, we consider the narrowed gate case. If the gate interval is decreased by $b$ from each side, we can conclude from Fig.~\ref{misalignment}(b) that $p_{\rm xtalk}=\mathrm{E}\{A\}$, where
\begin{equation}
A=\left\{
\begin{array}{cc}
 \eta^{\prime}\frac{2\mu (N-1)}{N^3T_p }(|\tau _{k}|-b-\delta)& |\tau _{k}|\geq b+\delta\\
0 & |\tau _{k}|\leq b+\delta
\end{array}
\right.
\end{equation}
Hence, $p_{\rm xtalk}$ can be written as
\begin{equation}
p_{\rm xtalk}=\eta^{\prime}\frac{2\mu (N-1)}{N^3 T_p}p(|\tau _{k}|\geq b+\delta)(\mathrm {E}\{|\tau _{k}|||\tau _{k}|\geq b+\delta \}-(b+\delta)).
\end{equation}
The loss factor $\eta_g$ is also obtained by $\mathrm{E}\{B\}$, where $B$ is given by
\begin{equation}
B=\left\{
\begin{array}{c}
\frac{1}{N}-\frac{2b}{NT_p}-\frac{1}{NT_p}(1-(\frac{N-1}{N})^2)((|\tau _{k}|-(b+\delta))\;\;\; |\tau _{k}|\geq b+\delta\\
\frac{1}{N}-\frac{|\tau _{k}|+b-\delta}{NT_p}\;\;\;\;\;\;\;\;\;\; \;\;\;\;\;\;\;\;\;\;\;\;\;\;\;\;\;\;\;\;\;\;\; |b-\delta|\leq |\tau _{k}|\leq b+\delta\\
\frac{1}{N}-\frac{2(b-\delta)}{NT_p}u(b-\delta) \;\;\;\;\;\;\;\;\;\;\;\;\;\;\;\;\;\;\;\;\;\;\;\;\;\;\;\;\;\;\; 0\leq |\tau _{k}|\leq |b-\delta|
\end{array}
\right.
\end{equation}
where $u(.)$ is the step function. Hence, $\eta_g$ can be expressed as
\begin{eqnarray}
\eta_g=\frac{1}{N}-\{(\frac{2b}{NT_p}-\frac{(b+\delta)}{NT_p}(1-(\frac{N-1}{N})^2))p(|\tau _{k}|\geq b+\delta)+ \nonumber\\
\frac{1}{NT_p}(1-(\frac{N-1}{N})^2)p(|\tau _{k}|\geq b+\delta)(\mathrm {E}\{|\tau _{k}|||\tau _{k}|\geq b+\delta\})+\nonumber\\
\frac{(b-\delta)}{NT_p}p(|b-\delta|\leq |\tau _{k}|\leq b+\delta))+
\frac{1}{NT_p}(p(|\tau _{k}|\geq |b-\delta|)\nonumber\\
 (\mathrm{E} \{|\tau _{k}|||\tau _{k}|\geq |b-\delta|\})-
p(|\tau _{k}|\geq b+\delta)(\mathrm{E} \{|\tau _{k}|||\tau _{k}|\geq b+\delta\}))\nonumber\\
-\frac{2(b-\delta)}{NT_p}p(|\tau _{k}|< b-\delta)u(b-\delta)\}\;\;\;\;\;\;\;\;\;\;\;\;\;\;\;\;\;\;\;\;\;\;\;\;\;\;\;\;\;\;\;\;\;
\label{loss_g_S_II}
\end{eqnarray}

Next, we discuss the time misalignment issue in Scheme II with an active OFDM decoder. Figure~\ref{misalignment}(c) shows the pulse series of the $k^{th}$ tributary in the presence of time misalignment $\tau _{k}$, $0<|\tau _{k}|<T_c$, with respect to the switching time. Due to the elimination of the loss incurred by passive serial to parallel conversion, equations (\ref{d_SchemeII_t1}) and (\ref{d_SchemeII_t}) are modified by a multiplicative factor $\sqrt{N}$. Hence, the parameters $p_{\rm xtalk}$ and $\eta_g$ are, respectively, given by (\ref{P_xtalk_II}) and (\ref{eta_g_s_2_p}), multiplied by $N$.
}
\bibliographystyle{IEEEtran}

\vspace{-.6in}
\begin{IEEEbiographynophoto}{Sima Bahrani}
received B.Sc. and M.Sc. degrees in Electrical Engineering from Shiraz University, Shiraz, Iran. From 2012, she is with Optical Networks Research Laboratory (ONRL), at Sharif University of Technology, Tehran, Iran, where she is currently pursuing her Ph.D. degree in Electrical Engineering. Her research interests include all-optical OFDM, Quantum communications, and Quantum cryptography. 
\end{IEEEbiographynophoto}

\vspace{-.5in}
\begin{IEEEbiographynophoto}{Mohsen Razavi}
 received his B.Sc. and M.Sc. degrees
(with honors) in Electrical Engineering from
Sharif University of Technology, Tehran, Iran, in
1998 and 2000, respectively. From August 1999 to
June 2001, he was a member of research staff at the
Iran Telecommunications Research Center, Tehran,
Iran, working on all-optical CDMA networks. He joined the Research Laboratory of
Electronics, at the Massachusetts Institute of Technology
(MIT), in 2001 to pursue his Ph.D. degree in
Electrical Engineering and Computer Science, which he completed in 2006.
He continued his work at MIT as a Post-doctoral Associate during Fall 2006,
before joining the Institute for Quantum Computing at the University of
Waterloo as a Post-doctoral Fellow in January 2007. He is currently an Associate Professor at the School of Electronic and Electrical Engineering at the University of Leeds. His research interests include a variety of problems
in quantum cryptography, quantum optics, and quantum communications networks.

\end{IEEEbiographynophoto}

\vspace{-.5in}
\begin{IEEEbiographynophoto}{Jawad A. Salehi}
(M’84–SM’07–F’10) received his B.S. degree from the University of California,
Irvine, in 1979, and the M.S. and Ph.D. degrees from the University of
Southern California, Los Angeles, in 1980 and 1984, respectively, all in electrical
engineering. He is currently a Full Professor with the Optical Networks Research Laboratory
(ONRL), Department of Electrical Engineering, Sharif University of
Technology, Tehran, Iran, where he is also the Co-founder of the Advanced
Communications Research Institute (ACRI). From 1981 to 1984, he was a
full-time Research Assistant at the Communication Science Institute, University
of Southern California. From 1984 to 1993, he was a Member of Technical
Staff of the Applied Research Area, Bell Communications Research (Bellcore),
Morristown, NJ. During 1990, he was with the Laboratory of Information and
Decision Systems, Massachusetts Institute of Technology (MIT), Cambridge,
as a Visiting Research Scientist. From 1999 to 2001, he was the Head of the
Mobile Communications Systems Group and the Co-director of the Advanced
and Wideband Code-Division Multiple Access (CDMA) Laboratory, Iran
Telecom Research Center (ITRC), Tehran. From 2003 to 2006, he was the
Director of the National Center of Excellence in Communications Science,
Department of Electrical Engineering, SUT. He holds 12 U.S. patents on optical
CDMA. His current research interests include optical multiaccess networks,
optical orthogonal codes (OOC), fiber-optic CDMA, femtosecond or ultrashort
light pulse CDMA, spread-time CDMA, holographic CDMA, wireless indoor
optical CDMA, all-optical synchronization, and applications of erbium-doped
fiber amplifiers in optical systems.
Prof. Salehi has been an Associate Editor for Optical CDMA of the IEEE TRANSACTIONS ON COMMUNICATIONS since May 2001. In September 2005, he was elected as the Interim Chair of the IEEE Iran Section. He was the recipient of several awards including Bellcore’s Award of Excellence, the Nationwide Outstanding Research Award from the Ministry of Science, Research, and Technology in 2003, and the Nation’s Highly Cited Researcher Award in 2004. In 2007, he received the Khwarazmi International Prize, first rank, in fundamental research and also the Outstanding Inventor Award (Gold medal) from the World Intellectual Property Organization (WIPO), Geneva, Switzerland. In 2010, he was promoted to IEEE Fellow  for contributions to the “fundamental principles in optical code division multiple access.” He is among the 250 pre-eminent and most influential researchers worldwide in the Institute for Scientific Information.
\end{IEEEbiographynophoto}

\end{document}